\documentclass[aps,prl,reprint,superscriptaddress,longbibliography,nobalancelastpage]{revtex4-2}

\usepackage{amsmath,amssymb}
\usepackage{graphicx}
\usepackage{xcolor}
\usepackage{mathrsfs}
\usepackage{stmaryrd}
\usepackage{hyperref}
\usepackage{enumitem}
\usepackage{accents}

\definecolor{linkcolor}{HTML}{223096}
\hypersetup{colorlinks,allcolors=linkcolor}
\bibpunct{\textcolor{linkcolor}{[}}{\textcolor{linkcolor}{]}}{\textcolor{linkcolor}{,}}{n}{}{;}
\renewcommand{\eqref}[1]{\hyperref[#1]{(\ref*{#1})}}

\newcommand{\figref}[2]{[Fig.~\hyperref[#1]{\ref*{#1}(#2)}]}
\newcommand{\figrefi}[2]{[Fig.~\hyperref[#1]{\ref*{#1}(#2)}, inset]}
\newcommand{\textfigref}[2]{Fig.~\hyperref[#1]{\ref*{#1}(#2)}}
\newcommand{\textfigureref}[2]{Figure~\hyperref[#1]{\ref*{#1}(#2)}}
\newcommand{\wholefigref}[1]{(Fig.~\ref{#1})}

\newcommand{\figrefp}[2]{\hyperref[#1]{\ref*{#1}(#2)}}

\renewcommand{\leq}{\leqslant}

\begin{document}

\title{Buckling by disordered growth}
\author{Rahul G. Ramachandran}
\affiliation{Max Planck Institute for the Physics of Complex Systems, N\"othnitzer Stra\ss e 38, 01187 Dresden, Germany}
\affiliation{Center for Systems Biology Dresden, Pfotenhauerstra\ss e 108, 01307 Dresden, Germany}
\author{Ricard Alert}
\altaffiliation{ralert@pks.mpg.de, haas@pks.mpg.de.}
\affiliation{Max Planck Institute for the Physics of Complex Systems, N\"othnitzer Stra\ss e 38, 01187 Dresden, Germany}
\affiliation{Center for Systems Biology Dresden, Pfotenhauerstra\ss e 108, 01307 Dresden, Germany}
\affiliation{Cluster of Excellence Physics of Life, TU Dresden, 01062 Dresden, Germany}
\author{Pierre A. Haas}
\altaffiliation{ralert@pks.mpg.de, haas@pks.mpg.de.}
\affiliation{Max Planck Institute for the Physics of Complex Systems, N\"othnitzer Stra\ss e 38, 01187 Dresden, Germany}
\affiliation{Center for Systems Biology Dresden, Pfotenhauerstra\ss e 108, 01307 Dresden, Germany}
\affiliation{\smash{Max Planck Institute of Molecular Cell Biology and Genetics, Pfotenhauerstra\ss e 108, 01307 Dresden, Germany}}
\date{\today}%
\begin{abstract}
Buckling instabilities driven by tissue growth underpin key developmental events such as the folding of the brain. Tissue growth is disordered due to cell-to-cell variability, but the effects of this variability on buckling are unknown. Here, we analyse what is perhaps the simplest setup of this problem: the buckling of an elastic rod with fixed ends driven by spatially varying growth. Combining analytical calculations for simple growth fields and numerical sampling of random growth fields, we show that variability can increase as well as decrease the growth threshold for buckling, even when growth variability does not cause any residual stresses. For random growth, we find that the shift of the buckling threshold correlates with spatial moments of the growth field. Our results imply that biological systems can either trigger or avoid buckling by exploiting the spatial arrangement of growth variability.
\end{abstract}

\maketitle

Mechanical instabilities can drive the development of bacterial biofilms, eukaryotic tissues, and organisms: Examples include the formation of biofilm wrinkles~\cite{yan19,fei20,cont20}, the gyrations of the brain~\cite{richman75,tallinen13,goriely15,tallinen16,balbi20}, the villi in the gut~\cite{hannezo11,savin11,shyer13,balbi15,gill24}\linebreak and the folding of the frill of the lizard \emph{Chlamydosaurus}~\cite{montandon19,haas19i}, which can be understood in terms of mechanical instabilities~\cite{nelson16} of buckling, wrinkling, and curtaining~\cite{cerda04}, while a new instability of ``buckling without bending'' has been associated with the formation of cerebellar folds~\cite{lejeune16,engstrom18,holland18,lawton19}.

These mechanical instabilities must therefore, just as development is robust at the tissue scale~\cite{vondassow07,cooper08,hong16,haas18a,hong18,yevick19,martin21,fruleux24}, be robust against the large amounts of cell-scale variability almost synonymous with biology. Remarkably, spatiotemporal growth variability of \emph{Arabidopsis} sepals is even necessary for ``correct'' sepal shapes~\cite{hong16,fruleux24}. Such variability can have a large effect on these morphogenetic instabilities: For example, brain tissue microstructure is highly heterogeneous~\cite{budday17}, which significantly influences the folding instabilities driving gyrification~\cite{budday18}. However, the physical mechanisms explaining the effect of variability on such mechanical instabilities that could resolve the resulting conundrum of cell-scale variability and tissue-scale robustness have remained largely unexplored. 

\begin{figure}[hb!]
\includegraphics{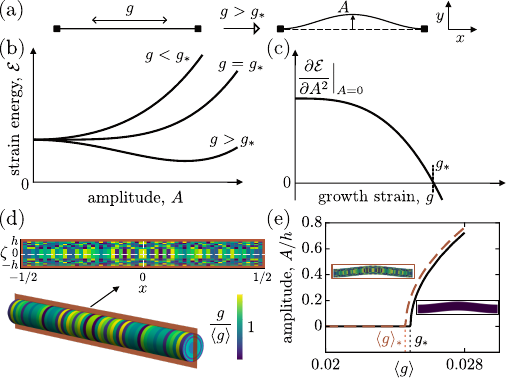}
\caption{\label{fig1}Buckling by disordered growth. (a)~Buckling of a growing elastic rod with clamped ends: if the axial growth $g$ exceeds the threshold $g_\ast$, the rod buckles with amplitude $A$. (b)~Elastic energy $\mathcal{E}$ against amplitude $A$ for $g<g_\ast$, $g=g_\ast$, $g>g_\ast$. If $g\leq g_\ast$, ${A=0}$ minimises~$\mathcal{E}$; for $g>g_\ast$, some $A>0$ minimises $\mathcal{E}$ and the rod buckles. (c)~Plot of $\partial\mathcal{E}/\partial A^2$ at $A=0$ against $g$, vanishing at $g_\ast$. (d)~Example of spatially varying growth $g$, plotted relative to the spatially-averaged mean growth $\langle g\rangle$. The growth field in the cross-section, with axial and radial coordinates $x\in[-1/2,1/2]$ and $\zeta\in[0,h]$, defines an axially and azimuthally symmetric growth field in the rod. (e)~Numerical buckling amplitude $A$ against mean growth $\langle g\rangle$. The buckling threshold $\langle g\rangle_\ast$ for the growth field in panel (d) (dashed line) is lower than the threshold $g_\ast$ for uniform growth (solid line).}
\end{figure}

Here, we therefore analyze a minimal mechanical model of spatial variability in mechanical instabilities by studying the buckling of a rod with disordered growth. Our starting point is a variant of the classical Euler buckling problem~\cite{euler,landaulifshitz}: An elastic rod of unit length, circular cross-section of radius $h$, and with clamped ends grows along its length. This axial growth, $g$, increases lengths locally by a factor $1+g$. When $g$ exceeds a critical growth $g_\ast$, the rod buckles out of its straight configuration with amplitude $A$ because the energetic cost of bending falls below that of compressing the rod further~\figref{fig1}{a}. We assume a buckled shape $y(x)=A\cos^2{\pi x}$ with ${A\ll 1}$, where ${x\in[-1/2,1/2]}$ is the coordinate along the undeformed rod midline. The elastic energy density of the rod is~\footnote{See Supplemental Material at [url to be inserted], which includes Refs.~\cite{moulton20,moulton13,green74,parker84,audoly21,steigmann13,haas21,docarmo,goriely,ambrosi19,rodriguez94,ogden,holzapfel2000nonlinear,DealIITutorial2012,pelteret2016b,rfs}, for (i)~a derivation of the elastic energy density $e$ of a uniformly growing rod, (ii)~a more formal derivation of the corresponding leading-order buckling threshold $g_\ast$, (iii) details of the numerical implementation of the finite-element simulations and of the random sampling of growth fields, (iv) details of the calculations leading to the approximation for the buckling threshold of a rod with growth islands.} ${e=C\bigl(E^2+h^2K^2/4\bigr)}$, where $C$ is a material parameter, $\smash{E=\sqrt{1+y'(x)^2}-(1+g)\approx y'(x)^2/2-g}$ is the midline strain (i.e., the difference of its actual and preferred, grown lengths), and ${K=y''(x)[1+y'(x)^2]^{-3/2}\approx y''(x)}$ is the curvature of this midline.\nocite{moulton20,moulton13,green74,parker84,audoly21,steigmann13,haas21,docarmo,goriely,ambrosi19,rodriguez94,ogden,holzapfel2000nonlinear,DealIITutorial2012,pelteret2016b,rfs} The energy of the rod, ${\mathcal{E}\approx C\pi h^2\left\{g^2\!+\!A^2[\pi^2(h^2\pi^2\!-\!g)/2]\right\}}$, is then obtained by integrating $e$ along its midline. At the critical\linebreak growth $g_\ast=h^2\pi^2$, $\partial\mathcal{E}/\partial A^2$ changes sign at ${A=0}$, and a buckled solution ($A\neq 0$) becomes favourable \figref{fig1}{b),(c}. The buckling amplitude $A$ for $g>g_\ast$ is set by nonlinearities beyond this calculation~\cite{Note1}.

In this work, we add quenched disorder to this picture by introducing spatial growth variability~\figref{fig1}{d),(e}, replacing the uniform growth $g$ with $g(x,\zeta)$, where ${x\in[-1/2,1/2]}$ is still the coordinate along the undeformed rod midline, and $\zeta\in[0,h]$ is the polar radius of its cross-section. We avoid twisting or asymmetric buckling of the rod by restricting to axially and azimuthally symmetric variability~\figref{fig1}{d}, i.e., ${g(x,\zeta,\phi)=g(x,\zeta)=g(-x,\zeta)}$. Combining finite-element simulations of the buckling of a growing elastic rod implemented within the deal.II library~\cite{Note1,dealII,DealIITutorial2012} and exact calculations, we analyse the effect of this variability on the buckling threshold. We will denote by $\langle g\rangle$ the (spatially averaged) mean growth and by $\langle g\rangle_\ast$ its value at the buckling threshold.

\begin{figure*}
\centering\includegraphics{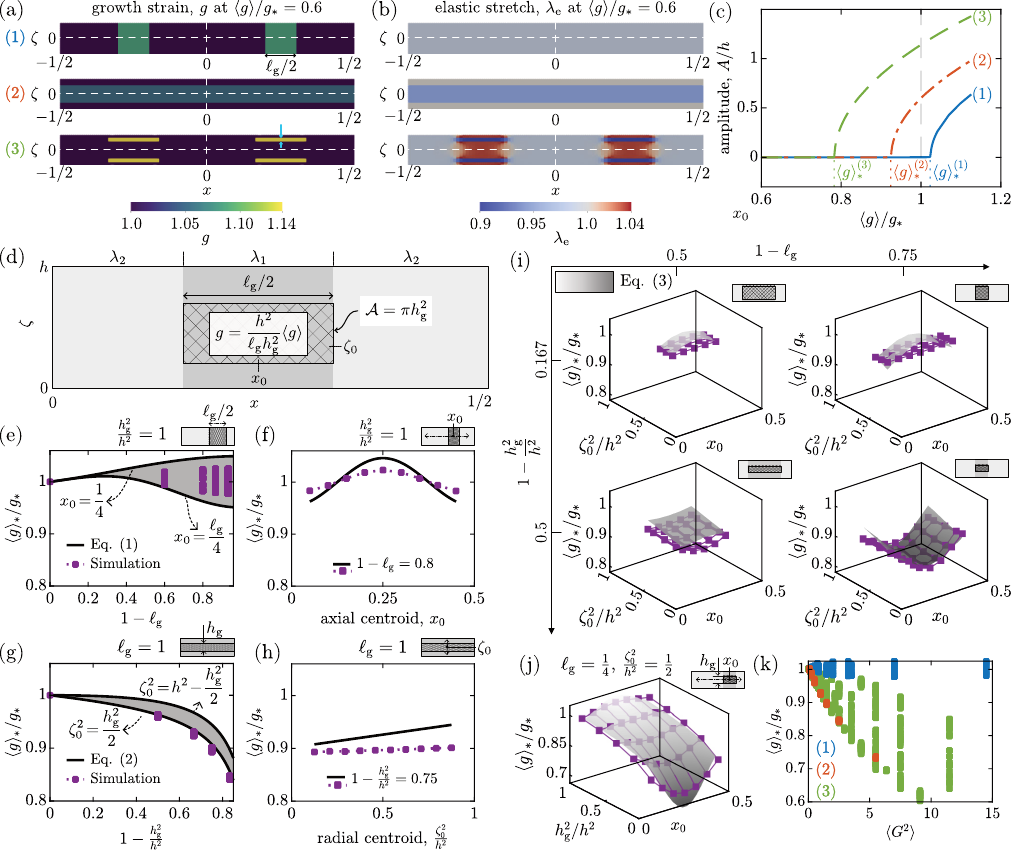}
\caption{Buckling with ``growth islands''. (a)~Cases of growth islands: (1)~axially inhomogeneous, radially homogeneous growth, with two symmetric axial segments of the rod spanning its full radius; (2) axially homogeneous, radially inhomogeneous growth, spanning part of the rod cross-section; (3) a general growth island consisting of two symmetric annular segments of the rod. The growth field $g$ is plotted at fixed mean growth $\langle g\rangle=0.6\,g_\ast$, lower than the buckling threshold in any of these examples. (b) Plot of the corresponding  elastic axial stretch field $\lambda_{\text{e}}$ from numerical calculations. The stretch fields are piecewise uniform in cases (1) and (2), but more complex in case (3). (c)~Numerical buckling amplitude against mean growth $\langle g\rangle$, for each of the three examples in panel~(a). The buckling threshold $\langle g\rangle_\ast$ differs from the uniform threshold $g_\ast$ in all three cases. (d)~Geometry of a growth island in a cross-section of the half-rod $0\leq x\leq 1/2$. Each growth island has axial extent $\ell_{\text{g}}/2$ and cross-sectional area $\mathcal{A}=\pi h_{\text{g}}^2$, so the growth in the growth island is $g=\langle g\rangle h^2/\ell_{\text{g}}h_{\text{g}}^2$, where $\langle g\rangle$ is the mean growth across the rod. The growth islands are centred axially at $x=\pm x_0$ and their radial centroid is at $\zeta=\zeta_0$. In calculations estimating the buckling threshold~\cite{Note1}, we assume that the deformed rod stretches piecewise uniformly, with stretches $\lambda_1$ in cross-sections containing the growth island and $\lambda_2$ elsewhere. (e)~Plot of the buckling threshold $\langle g\rangle_\ast$ against $\ell_{\text{g}}$ for axially inhomogeneous, radially homogeneous growth islands [case (1), illustrated by the inset analogous to panel (d) at the top of the plot], showing simulation results for different $x_0$ (marks) and the range estimated from Eq.~\eqref{eq:e1} (shaded area). (f)~Corresponding plot of $\langle g\rangle_\ast$ against $x_0$ at fixed $\ell_{\text{g}}$. (g)~Similar plot of $\langle g\rangle_\ast$ against $h_{\text{g}}$ for axially homogeneous, radially inhomogeneous growth islands [case~(2)], showing simulation results for different $\zeta_0$ (marks) and the estimated range from Eq.~\eqref{eq:e2} (shaded area). (h)~Corresponding plot of $\langle g\rangle_\ast$ against $\zeta_0$ at fixed~$h_{\text{g}}$. (i)~Buckling threshold for general growth islands [case (3), illustrated by the insets at the top of the plots]: Plots of~$\langle g\rangle_\ast$ against $x_0$ and $\zeta_0$ for different $\ell_{\text{g}}$ and $h_{\text{g}}$, showing numerical results (marks) and estimates from Eq.~\eqref{eq:e3} (shaded surfaces). A flipping transition of the buckling threshold behaviour, discussed in the text, is seen as $h_{\mathrm{g}}$ is reduced. (j)~Corresponding plot of $\langle g\rangle_\ast$ at fixed~$\ell_{\text{g}}$ and $\zeta_0$ and for varying $h_{\text{g}}$ and $x_0$, as illustrated in the inset, again showing the flipping transition. (k)~Plot of $\langle g\rangle_\ast$ against the variance $\langle G^2\rangle$ of the growth disorder $G(x,\zeta)=g(x,\zeta)/\langle g\rangle-1$, for different growth islands (in all three cases), showing the correlation between the buckling threshold and this variance.\label{fig2}}
\end{figure*}

We begin by studying ``growth islands'', where part of the rod grows uniformly, while the rest does not grow~\figref{fig2}{a}. This includes the subcases in which the growth islands are (1) two symmetric axial segments spanning the whole thickness of the rod, and (2) an annulus of the cross-section of the rod spanning its entire length. In the most general case, the growth islands are (3)~two symmetric annular segments. In case (1), our numerical calculations give a uniform elastic stretch of the compressed rod before buckling. The stretch remains piecewise uniform along the rod in case~(2), while, in case~(3), the elastic stretch field is more complex~\figref{fig2}{b}. In all three cases, the buckling threshold changes compared to the case of uniform growth, $\langle g\rangle_\ast\neq g_\ast$. This is hardly surprising in cases (2) and~(3) because the growth field is incompatible with the geometry of the rod. This incompatibility produces residual stresses which might be expected to affect the buckling threshold. The surprise, however, is in case (1), in which the growth field is compatible with the rod geometry. Hence, the rod is unstressed if its ends are not fixed, and it has the same grown length as a uniformly growing rod, and yet the buckling thresholds differ! 

\textfigureref{fig2}{c} also illustrates that spatial variability can result in an increase and a decrease of the buckling threshold. In particular, increased buckling thresholds disprove the naive explanation that the geometric incompatibilities resulting from growth disorder generate additional stresses that favour buckling.

To explain these observations, we therefore derive, in the Supplemental Material~\cite{Note1}, an analytical estimate of the buckling threshold for growth islands of length $\ell_{\text{g}}$ and cross-sectional area $\smash{\pi h_{\text{g}}^2}$, centred at $x=\pm x_0$ and $\zeta=\zeta_0$ \figref{fig2}{d}. In this calculation extending that in the introduction, we assume the stretch of the compressed rod to be piecewise uniform, consistently with the numerical observations in \textfigref{fig2}{b} for cases (1) and (2).

In case (1), in which the growth island spans the thickness of the rod ($h_{\text{g}}=h$), we obtain~\cite{Note1}
\begin{align}
\dfrac{\langle g\rangle_\ast}{g_\ast}\approx 1-\dfrac{2\pi h^2}{\ell_{\text{g}}^2}\cos{\bigl(4\pi x_0\bigr)}\sin{\bigl(\pi\ell_{\text{g}}\bigr)}.\label{eq:e1}
\end{align}
This estimate is \emph{not} an asymptotic approximation of the buckling threshold~\footnote{Because the analytical estimates suggest that the effect of variability is a subleading correction to the uniform buckling threshold, deriving asymptotically correct expressions for buckling thresholds, involving in particular the spatial moments of the growth field, would (likely) require not only obtaining the stress in the deformed cross-section of the rod, but would also involve the corrections, from an asymptotic expansion of three-dimensional elasticity, to the rod theory underlying this calculation.}. Still, it captures qualitatively the range of buckling thresholds for different $\ell_{\text{g}}$ and the variation of the buckling threshold with $x_0$ in numerical simulations~\figref{fig2}{e),(f}. In particular, from \textfigref{fig2}{f} or Eq.~\eqref{eq:e1}, the buckling threshold is maximal, at fixed $\ell_{\text{g}}$, if the growth islands are in the middle of the two rod halves ($x_0=\pm 1/4$), and minimal when they are at the ends or in the middle of the rod (${\pm x_0=\ell_{\text{g}}/4,1-\ell_{\text{g}}/4}$)~\figref{fig2}{d}.

Similarly, in case (2), in which the growth island spans the length of the rod ($\ell_{\text{g}}=1$), we estimate~\cite{Note1}
\begin{align}
\dfrac{\langle g\rangle_\ast}{g_\ast}\approx 1+\pi^2h^2\left[2\left(\dfrac{\zeta_0}{h}\right)^2-\left(\dfrac{h_{\text{g}}}{h}\right)^{-2}\right],\label{eq:e2}
\end{align}
which is qualitatively consistent with the numerical range of the buckling thresholds for different $h_{\text{g}}$ and the variation of the buckling threshold with $\zeta_0$ seen in numerical simulations~\figref{fig2}{g),(h}. 

For the general growth islands of case (3), we observe numerically that the buckling threshold is largest if the growth is in the middle of the two rod halves (${x_0=\pm 1/4}$) if $h_{\text{g}}$ is sufficiently large~\figref{fig2}{i}, for all $\zeta_0$ and $\ell_{\text{g}}$, in agreement with the observations and calculations for case~(1) in \textfigref{fig2}{f}. Intriguingly, this trend flips for smaller $h_{\text{g}}$, i.e., for more localised growth~\figref{fig2}{i}, in which case the threshold is largest when the growth islands are at the ends or in the middle of the rod (${\pm x_0=\ell_{\text{g}}/4,1-\ell_{\text{g}}/4}$). The elastic stretch field in the rod is more complex in case (3)~\figref{fig2}{b} than the piecewise uniform stretch field that we assume in calculations. Surprisingly, however, the estimate~\cite{Note1}
\begin{widetext}

\vspace{-12pt}
\begin{align}
\dfrac{\langle g\rangle_\ast}{g_\ast}\approx 1+\dfrac{\pi h^2}{\ell_{\text{g}}^2}\left\{\pi\ell_{\text{g}}\left[1\!-\!\ell_{\text{g}}\!+\!2\ell_{\text{g}}\left(\dfrac{\zeta_0}{h}\right)^2\!-\!\left(\dfrac{h_{\text{g}}}{h}\right)^{-2}\right]-\left[1\!+\!3\ell_{\text{g}}\!-\!2\ell_{\text{g}}\left(\dfrac{\zeta_0}{h}\right)^2\!-\!\left(\dfrac{h_{\text{g}}}{h}\right)^{-2}\right]\cos{\bigl(4\pi x_0\bigr)}\sin{\bigl(\pi\ell_{\text{g}}\bigr)}\right\},\label{eq:e3}
\end{align}

\vspace{-5pt}
\end{widetext}
of which Eqs.~\eqref{eq:e1} and \eqref{eq:e2} are special cases, still captures the numerically observed flipping behaviour qualitatively~\figref{fig2}{i}. In particular, Eq.~\eqref{eq:e3} predicts flipping when the second term in square brackets changes sign, i.e., for $h_{\text{g}}/h<\{1+\ell_{\text{g}}+2\ell_{\text{g}}[1-(\zeta_0/h)^2]\}^{-1/2}$. This transition can also be visualised and reproduced qualitatively by Eq.~\eqref{eq:e3} at fixed $\ell_{\text{g}}$ and $\zeta_0$ as $h_{\text{g}}$ is varied~\figref{fig2}{j}. 

It is natural to seek a statistical description of this rich mechanical effect of spatial variability on the buckling threshold. We therefore ask how the buckling threshold relates to the statistical moments of the growth disorder $G(x,\zeta)$, where ${g(x,\zeta)=[1+G(x,\zeta)]\langle g\rangle}$. By definition, $\langle G\rangle=0$, so the variance $\langle G^2\rangle$ is the first non-trivial such moment. The small changes of $\langle g\rangle_\ast$ in case~(1) do not correlate strongly with $\langle G^2\rangle$, but the larger changes in case~(2) do correlate well with this variance~\figref{fig2}{k}. This correlation is also apparent for general growth islands~[case~(3)], although the spread of buckling thresholds is much higher~\figref{fig2}{k}.

\begin{figure}[b]
\centering\includegraphics{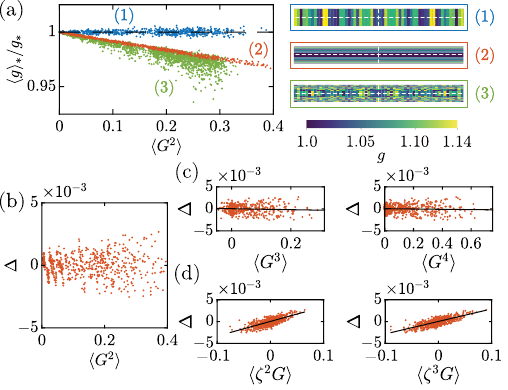}
\caption{``Mechanical statistics'' of buckling with disorder. (a)~Plot of the buckling threshold $\langle g\rangle_\ast$ against the variance $\langle G^2\rangle$ of the growth disorder for randomly sampled growth fields~\cite{Note1} in three cases (insets): (1) axially inhomogeneous, radially homogeneous growth, (2) axially homogeneous, radially inhomogeneous growth, and (3) general, axially and radially inhomogeneous growth, by analogy with the three classes of growth islands~\figref{fig2}{a}. (b) Plot of the difference $\smash{\Delta=\langle g\rangle_\ast/g_\ast-\smash{\bigl\langle \langle g\rangle_\ast/g_\ast\bigr\rangle}_{\smash{\langle G^2\rangle}}}$ between the buckling threshold and its average at fixed $\langle G^2\rangle$ against $\langle G^2\rangle$, for radially inhomogeneous growth [case (2)]. (c)~For these growth fields in case (2), $\Delta$ does not correlate well with the higher moments $\langle G^3\rangle,\langle G^4\rangle,\dots$ of the growth disorder $G$. (d) For these growth fields in case~(2), $\Delta$ does correlate with the spatial moments $\langle \zeta G\rangle,\langle\zeta^2G\rangle,\dots$. \label{fig3}}
\end{figure}

We now extend these results beyond growth islands by performing ``mechanical statistics'': We sample random growth fields from a uniform probability distribution~\cite{Note1} and obtain the resulting buckling threshold~\figref{fig3}{a} in finite-element simulations~\cite{Note1}. We distinguish again between (1) axially inhomogeneous, radially homogeneous and (2) axially homogeneous, radially inhomogeneous growth fields as particular cases of (3) general, axially and radially inhomogeneous growth fields. Once again, the buckling threshold does not vary much in case (1), but varies more strongly in cases (2) and (3), and this variation again correlates well with the variance $\langle G^2\rangle$ of the growth disorder~\figref{fig3}{a}, especially so in case (2). 

One might expect this correlation of $\langle g\rangle_\ast/g_\ast$ and $\langle G^2\rangle$: Naively, buckling occurs when the local growth $g(x,\zeta)$ reaches the critical growth for homogeneous buckling, $g_\ast$, i.e. when $\langle g\rangle/g_\ast=[1+G(x,\zeta)]^{-1}$ locally. Averaging spatially, this suggests $\langle g\rangle_\ast/g_\ast\approx 1+\langle G^2\rangle$, where we have used that $\langle G\rangle=0$ by definition and expanded at small disorder. This predicts that $\langle g\rangle_\ast/g_\ast$ increases with $\langle G^2\rangle$, rather than decreasing with $\langle G^2\rangle$ as observed in cases (2) and~(3)~\figref{fig3}{a}. This failure of the local buckling argument emphasises the role of the global mechanics of the rod. Moreover, this naive argument suggests that those variations of the buckling threshold that are not explained by the variance $\langle G^2\rangle$ are related to higher moments $\langle G^3\rangle,\langle G^4\rangle,\dots$ of the growth disorder. However, focusing now on case (2), and introducing the difference ${\Delta=\langle g\rangle_\ast/g_\ast-\smash{\bigl\langle \langle g\rangle_\ast/g_\ast\bigr\rangle}_{\smash{\langle G^2\rangle}}}$ between the buckling threshold and the average buckling threshold at fixed variance of the growth disorder~\figref{fig3}{b}, we find that $\Delta$ does not correlate well with these higher moments~\figref{fig3}{c}. Strikingly, we do find that $\Delta$ correlates better with the spatial moments $\langle \zeta G\rangle,\langle\zeta^2 G\rangle,\dots$ of the growth disorder~\figref{fig3}{d}. This shows that disorder shifts the buckling threshold not only via its variance, but also via these spatial moments. In turn, this implies that the probability distribution of buckling thresholds depends not only on the probability distribution from which the growth variability is sampled, but also on the distribution of the spatial moments that emerges from it at the scale of the rod.

In summary, we have shown how spatial variability in growth shifts the buckling transition of an elastic rod. To understand these effects, we have proposed analytical estimates for growth islands~\wholefigref{fig2}. For random growth fields, we found that the shift of the buckling threshold correlates with spatial moments of the growth field~\wholefigref{fig3}. Together with the probability distribution of the growth disorder, the spatial moments of the growth field thus control buckling and hence, more biologically, its robustness against microscopic, cell-scale disorder.

Our results have multiple additional implications for the robustness of buckling transitions in biological systems. First, we showed that variability can significantly decrease the buckling threshold, but we did not find any cases that would lead to a comparably large increase of the threshold~\figref{fig3}{a}, suggesting that variability favours morphogenetic instabilities. Next, case (1) in \textfigref{fig3}{a} might represent growth disorder in a cell monolayer. The buckling threshold varies but slightly in this case, which suggests that the buckling transition of cell monolayers is largely independent of in-plane disorder. Since the threshold varies both above and below that for uniform buckling, imposing the average growth required for uniform buckling is not sufficient to ensure buckling in the presence of disorder. Finally, the spatial moments that we have implicated in the buckling threshold~\figref{fig3}{d} stress that robustness depends not only on the magnitude of cell-to-cell variability but also on its spatial distribution in the tissue.

Meanwhile, performing the asymptotic calculations of buckling thresholds~\cite{Note2} that would confirm our analytical estimates and numerical results remains an open challenge. Our work also opens the door to interweaving the ideas of ``mechanical statistics'' that we have introduced here with the more established concepts of stochastic elasticity~\cite{staber15,staber17,mihai18,mihai19,mihai22}, which studies ensembles of elastic objects with \emph{uniform} material properties drawn from a probability distribution.

\begin{acknowledgments}
The authors gratefully acknowledge funding from the Max Planck Society.
\end{acknowledgments}

\bibliography{main}
\end{document}


\renewcommand{\theequation}{S\arabic{equation}}
\renewcommand{\thefigure}{S\arabic{figure}}

\title{Buckling by disordered growth\\ \textbullet{} Supplemental Material \textbullet{}}
\author{Rahul G. Ramachandran}
\affiliation{Max Planck Institute for the Physics of Complex Systems, N\"othnitzer Stra\ss e 38, 01187 Dresden, Germany}
\affiliation{Center for Systems Biology Dresden, Pfotenhauerstra\ss e 108, 01307 Dresden, Germany}
\author{Ricard Alert}
\affiliation{Max Planck Institute for the Physics of Complex Systems, N\"othnitzer Stra\ss e 38, 01187 Dresden, Germany}
\affiliation{Center for Systems Biology Dresden, Pfotenhauerstra\ss e 108, 01307 Dresden, Germany}
\affiliation{Cluster of Excellence Physics of Life, TU Dresden, 01062 Dresden, Germany}
\author{Pierre A. Haas}
\affiliation{Max Planck Institute for the Physics of Complex Systems, N\"othnitzer Stra\ss e 38, 01187 Dresden, Germany}
\affiliation{Center for Systems Biology Dresden, Pfotenhauerstra\ss e 108, 01307 Dresden, Germany}
\affiliation{\smash{Max Planck Institute of Molecular Cell Biology and Genetics, Pfotenhauerstra\ss e 108, 01307 Dresden, Germany}}

\maketitle

This Supplemental Material is divided into four parts. First, we derive the asymptotic energy density of a uniformly growing thin rod. We then provide a more formal derivation of the expression for the buckling threshold of the uniformly growing rod. Next, we give details of the numerical implementation of the finite-element simulations and the random sampling of growth fields. Finally, we provide the calculations for the approximate buckling threshold of a rod with growth islands.

\vspace{-8pt}
\section{Mechanics of a uniformly growing rod}
\vspace{-8pt}
In this section, we derive the energy of a uniformly growing, incompressible elastic rod with circular cross-section of radius $\varepsilon h$. This underlies the calculation of the buckling threshold of a uniformly growing rod in the main text and in the second part of this Supplemental Material, as well as the the analytical approximation for the buckling threshold of a rod with growth islands in the final part of this Supplemental Material.

Such a rod theory has previously been derived from three-dimensional elasticity~\cite{moulton20} or based on the Kirchhoff rod theory~\cite{moulton13}; theories for non-growing rods have also been derived from three-dimensional elasticity~\cite{green74,parker84,audoly21}. Here, we provide an alternative derivation  by asymptotic expansion, in the limit $\varepsilon\ll 1$, of three-dimensional nonlinear elasticity. This directly yields the expression of the energy density needed in our calculations. The derivation follows the ideas of the derivation of elastic shell theories in Refs.~\cite{steigmann13,haas21}, with the notation based on that of Ref.~\cite{haas21}. 

\vspace{-8pt}
\subsection{Rod geometry}
\vspace{-8pt}
Let $\vec{\rho}(s)$ be the undeformed midline of the rod, where $s$ is arclength, and let $\{\vec{t},\vec{n},\vec{b}\}$ be the corresponding Frenet basis~\cite{docarmo}. The position vector of a point in the rod is then
\begin{align}
\vec{r}(s,\zeta,\phi)=\vec{\rho}(s)+\varepsilon\zeta[\vec{b}(s)\cos{\phi}+\vec{n}(s)\sin{\phi}],
\end{align}
where $(\zeta,\phi)$ are polar coordinates for the circular cross-section of the rod, with $0\leq\zeta\leq h$. Using the Frenet--Serret equations~\cite{docarmo}, we then compute the partial derivatives
\begin{subequations}\label{eq:dundef}
\begin{align}
\dfrac{\partial\vec{r}}{\partial s}&=(1-\varepsilon\kappa\sin{\phi})\vec{t}-\varepsilon\zeta\tau(\vec{n}\cos{\phi}-\vec{b}\sin{\phi}),\nonumber\\
&=(1-\varepsilon\kappa\sin{\phi})\vec{t}+\varepsilon\zeta\tau\vec{B}\\
\dfrac{\partial\vec{r}}{\partial \zeta}&=\varepsilon(\vec{b}\cos{\phi}+\vec{n}\sin{\phi})=\varepsilon\vec{N},\\
\dfrac{\partial\vec{r}}{\partial \phi}&=\varepsilon\zeta(-\vec{b}\sin{\phi}+\vec{n}\cos{\phi})=-\varepsilon\zeta\vec{B},
\end{align}
\end{subequations}
where $\kappa$ and $\tau$ are the torsion and curvature of the undeformed rod, respectively, and where
\begin{align}
&\vec{N}=\vec{b}\cos{\phi}+\vec{n}\sin{\phi},&&\vec{B}=\vec{b}\sin{\phi}-\vec{n}\cos{\phi},
\end{align}
so that $\mathcal{B}=\{\vec{t},\vec{N},\vec{B}\}$ also defines a (right-handed) orthonormal basis, in which, by construction, $\vec{N}$ is the normal to the surface of the rod. The metric scale factors of the undeformed rod are therefore
\begin{align}
\chi_s&=\left[(1\!-\!\varepsilon\kappa\sin{\phi})^2+(\varepsilon\zeta\tau)^2\right]^{1/2},\;\chi_\zeta=\varepsilon,\;\chi_\phi=\varepsilon\zeta.\label{eq:scalef}
\end{align}
The deformed configuration of the rod is similarly characterised by its midline $\vec{\tilde{\rho}}(s)$, which has arclength $\tilde{s}(s)$. We denote by $\bigl\{\vec{\tilde{t}},\vec{\tilde{n}},\vec{\tilde{b}}\bigr\}$ its Frenet basis, and write the position vector of a point in the deformed rod as 
\begin{align}
\vec{\tilde{r}}(s,\zeta,\phi)&=\nonumber\vec{\tilde{\rho}}(s)+\tilde{\zeta}(s,\zeta,\phi)\bigl[\vec{\tilde{b}}(s)\cos{\tilde{\phi}(s,\zeta,\phi)}\nonumber\\
&\qquad+\vec{\tilde{n}}(s)\sin{\tilde{\phi}(s,\zeta,\phi)}\bigr]+\varepsilon\tilde{\varsigma}(s,\zeta,\phi)\vec{\tilde{t}}(s)\nonumber\\
&=\vec{\tilde{\rho}}(s)\!+\!\tilde{\zeta}(s,\zeta,\phi)\vec{\tilde{N}}(s,\phi)\!+\!\varepsilon\tilde{\varsigma}(s,\zeta,\phi)\vec{\tilde{t}}(s),
\end{align}
where $\tilde{\zeta}(s,\zeta,\phi)$ is the polar radial coordinate of the deformed cross-section of the rod, which may not be circular, and $\tilde{\phi}(s,\zeta,\phi)$ is the corresponding azimuthal coordinate, and where $\tilde{\varsigma}(s,\zeta,\phi)$ describes the deformation of that cross-section parallel to the rod midline. We have also introduced
\begin{align}
&\vec{\tilde{N}}=\vec{\tilde{b}}\cos{\tilde{\phi}}+\vec{\tilde{n}}\sin{\tilde{\phi}},&&\vec{\tilde{B}}=\vec{\tilde{b}}\sin{\tilde{\phi}}-\vec{\tilde{n}}\cos{\tilde{\phi}},\label{eq:BN}
\end{align}
to define a (right-handed) orthonormal basis ${\tilde{\mathcal{B}}=\bigl\{\vec{\tilde{t}},\vec{\tilde{N}},\vec{\tilde{B}}\bigr\}}$ for the deformed rod. We thus compute the partial derivatives
\begin{subequations}\label{eq:ddef}
\begin{align}
\dfrac{\partial\vec{\tilde{r}}}{\partial s}&=\left[\tilde{f}\bigl(1-\varepsilon\tilde{\kappa}\tilde{\zeta}\sin{\tilde{\phi}}\bigr)+\varepsilon\tilde{\varsigma}_{,s}\right]\vec{\tilde{t}}+\varepsilon\left(\tilde{\zeta}_{,s}+\tilde{f}\tilde{\varsigma}\tilde{\kappa}\sin{\tilde{\phi}}\right)\vec{\tilde{N}}\nonumber\\
&\qquad+\varepsilon\left(\tilde{f}\tilde{\zeta}\tilde{\tau}-\tilde{\zeta}\tilde{\phi}_{,s}-\tilde{f}\tilde{\varsigma}\tilde{\kappa}\cos{\tilde{\phi}}\right)\vec{\tilde{B}},\\
\dfrac{\partial\vec{\tilde{r}}}{\partial \zeta}&=\varepsilon\tilde{\varsigma}_{,\zeta}\vec{\tilde{t}}+\varepsilon\tilde{\zeta}_{,\zeta}\vec{\tilde{N}}-\varepsilon\tilde{\zeta}\tilde{\phi}_{,\zeta}\vec{\tilde{B}},\\
\dfrac{\partial\vec{\tilde{r}}}{\partial \phi}&=\varepsilon\tilde{\varsigma}_{,\phi}\vec{\tilde{t}}+\varepsilon\tilde{\zeta}_{,\phi}\vec{\tilde{N}}-\varepsilon\tilde{\zeta}\tilde{\phi}_{,\phi}\vec{\tilde{B}},
\end{align}
\end{subequations}
where $\tilde{f}=\mathrm{d}\tilde{s}/\mathrm{d}s$ is the stretch of the deformed rod midline, $\tilde{\kappa}$ and $\tilde{\tau}$ are its curvature and torsion, and wherein commata denote differentiation. By definition, the deformation gradient tensor for the deformation of the rod is~\cite{goriely}
\begin{subequations}
\begin{align}
\tens{\tilde{F}}=\operatorname{Grad}\vec{\tilde{r}}=\dfrac{1}{\chi_s^2}\dfrac{\partial\vec{\tilde{r}}}{\partial s}\otimes\dfrac{\partial\vec{r}}{\partial s}+\dfrac{1}{\chi_\zeta^2}\dfrac{\partial\vec{\tilde{r}}}{\partial\zeta}\otimes\dfrac{\partial\vec{r}}{\partial\zeta}+\dfrac{1}{\chi_\phi^2}\dfrac{\partial\vec{\tilde{r}}}{\partial\phi}\otimes\dfrac{\partial\vec{r}}{\partial\phi}.
\end{align}
Hence, from Eqs.~\eqref{eq:dundef}, \eqref{eq:scalef}, \eqref{eq:ddef}, we obtain an explicit expression for $\tens{\tilde{F}}$ with respect to $\tilde{\mathcal{B}}\otimes\mathcal{B}$,
\begin{widetext}
\begin{align}
\tens{\tilde{F}}=\left(\begin{array}{ccc}
\vspace{4pt}
\dfrac{\tilde{f}\bigl(1-\varepsilon\tilde{\kappa}\tilde{\zeta}\sin{\tilde{\phi}}\bigr)+\varepsilon\tilde{\varsigma}_{,s}}{(1-\varepsilon\kappa\sin{\phi})^2+(\varepsilon\zeta\tau)^2}(1-\varepsilon\kappa\sin{\phi})&\tilde{\varsigma}_{,\zeta}&\dfrac{\tilde{f}\bigl(1-\varepsilon\tilde{\kappa}\tilde{\zeta}\sin{\tilde{\phi}}\bigr)+\varepsilon\tilde{\varsigma}_{,s}}{(1-\varepsilon\kappa\sin{\phi})^2+(\varepsilon\zeta\tau)^2}\varepsilon\zeta\tau-\dfrac{\tilde{\varsigma}_{,\phi}}{\zeta}\\
\vspace{4pt}
\dfrac{\varepsilon\left(\tilde{\zeta}_{,s}+\tilde{f}\tilde{\varsigma}\tilde{\kappa}\sin{\tilde{\phi}}\right)}{(1-\varepsilon\kappa\sin{\phi})^2+(\varepsilon\zeta\tau)^2}(1-\varepsilon\kappa\sin{\phi})&\tilde{\zeta}_{,\zeta}&\dfrac{\varepsilon\left(\tilde{\zeta}_{,s}+\tilde{f}\tilde{\varsigma}\tilde{\kappa}\sin{\tilde{\phi}}\right)}{(1-\varepsilon\kappa\sin{\phi})^2+(\varepsilon\zeta\tau)^2}\varepsilon\zeta\tau-\dfrac{\tilde{\zeta}_{,\phi}}{\zeta}\\
\dfrac{\varepsilon\left(\tilde{f}\tilde{\zeta}\tilde{\tau}-\tilde{\zeta}\tilde{\phi}_{,s}-\tilde{f}\tilde{\varsigma}\tilde{\kappa}\cos{\tilde{\phi}}\right)}{(1-\varepsilon\kappa\sin{\phi})^2+(\varepsilon\zeta\tau)^2}(1-\varepsilon\kappa\sin{\phi})&-\tilde{\zeta}\tilde{\phi}_{,\zeta}&\dfrac{\varepsilon\left(\tilde{f}\tilde{\zeta}\tilde{\tau}-\tilde{\zeta}\tilde{\phi}_{,s}-\tilde{f}\tilde{\varsigma}\tilde{\kappa}\cos{\tilde{\phi}}\right)}{(1-\varepsilon\kappa\sin{\phi})^2+(\varepsilon\zeta\tau)^2}\varepsilon\zeta\tau+\dfrac{\tilde{\zeta}\tilde{\phi}_{,\phi}}{\zeta}
\end{array}\right).
\end{align}
\end{widetext}
\end{subequations}
\subsection{Rod elasticity}
We describe the elasticity of the rod within the framework of morphoelasticity~\cite{goriely,ambrosi19}, based on a multiplicative decomposition~\cite{rodriguez94} of the deformation gradient $\tens{\tilde{F}}$ into an intrinsic growth part $\tens{F_g}$ and an elastic part $\smash{\tens{F}=\tens{\tilde{F}}\bigl(\tens{F_g}\bigr)^{-1}}$. With incompressible neo--Hookean constitutive relations~\cite{goriely}, the deformed configuration of the rod minimises the elastic energy
\begin{align}
\mathcal{E}=\iiint_{\mathcal{V}_g}{e\,\mathrm{d}V_g},\quad\text{with }e=\dfrac{C}{2}\bigl[\operatorname{tr}\bigl(\tens{F}^\top\tens{F}\bigr)-3\bigr],\label{eq:E}
\end{align}
subject to the incompressibility constraint $\det{\tens{F}}=1$. Here, $C>0$ is a material parameter and the integration is over the grown virtual configuration $\mathcal{V}_g$ of the rod, with volume element $\mathrm{d}V_g=\bigl(\det{\tens{F_g}}\bigr)\,\mathrm{d}V$, where $\mathrm{d}V=\varepsilon^2\zeta\,\mathrm{d}s\,\mathrm{d}\zeta\,\mathrm{d}\phi$ is the volume element of the undeformed configuration.

The morphoelastic analogue of the first Piola--Kirchhoff stress tensor~\cite{goriely,ogden} is
\begin{align}
\tens{P}=C\bigl[\tens{F}\bigl(\tens{F_g}\bigr)^{-\top}-p\tens{\tilde{F}}^{-\top}\bigr],
\end{align}
where $p$ is the pressure that enforces incompressibility~\cite{haas21}. The Cauchy equation~\cite{goriely,ogden} then takes the form
\begin{subequations}\label{eq:cauchy}
\begin{align}
\operatorname{Div}\tens{P}^\top=\vec{0},
\end{align}
subject to the force-free condition $\tens{P}\cdot\vec{N}=\vec{0}$ on the surface $\zeta=h$ of the undeformed rod, and where the divergence is taken with respect to the undeformed configuration of the shell. Using its definition~\cite{goriely}, we find
\begin{align}
\operatorname{Div}\tens{P}^\top&=\dfrac{(1-\varepsilon\kappa\sin{\phi})\vec{t}+\varepsilon\zeta\tau\vec{B}}{(1-\varepsilon\kappa\sin{\phi})^2+(\varepsilon\zeta\tau)^2}\cdot\dfrac{\partial\tens{P}^\top}{\partial s}\nonumber\\
&\qquad+\dfrac{1}{\varepsilon\zeta}\left[\dfrac{\partial}{\partial \zeta}\bigl(\zeta\tens{P}\cdot\vec{N}\bigr)-\dfrac{\partial}{\partial \phi}\bigl(\tens{P}\cdot\vec{B}\bigr)\right].
\end{align}
\end{subequations}
\subsection{Asymptotic expansion}
We now specialise to the case of an axially growing rod that we are interested in here, and which corresponds to
\begin{align}
\tens{F_g}=\left(\begin{array}{ccc}
1+\varepsilon g&0&0\\
0&1&0\\
0&0&1
\end{array}\right),
\end{align}
where $g$ is the constant axial growth, and we will ignore torsion of the rod, $\tau=\tilde{\tau}=0$. We will further assume that deformations of the rod obey the small-strain scalings
\begin{align}
&\tilde{f}=1+\varepsilon g+\varepsilon E,&&\tilde{\kappa}=\kappa+K,\label{eq:EK}
\end{align}
which define the midline strain $E$ and curvature strain $K$ of the rod. The derivation of the rod theory now involves minimisation of the elastic energy~\eqref{eq:E} order by order, solving for the deformations
\begin{subequations}
\begin{align}
\tilde{\zeta}&=Z_0+\varepsilon Z_1+\varepsilon^2Z_2+O\bigl(\varepsilon^3\bigr),\\
\tilde{\phi}&=\Phi_0+\varepsilon \Phi_1+\varepsilon^2\Phi_2+O\bigl(\varepsilon^3\bigr),\\
\tilde{\varsigma}&=S_0+\varepsilon S_1+\varepsilon^2S_2+O\bigl(\varepsilon^3\bigr),
\end{align}
\end{subequations}
and the pressure $p=p_0+\varepsilon p_1+O\bigl(\varepsilon^2\bigr)$. These expansions in turn define an expansion $\smash{\tens{P}=\tens{P_0}+\varepsilon\tens{P_1}+O\bigl(\varepsilon^2\bigr)}$. 

In this minimisation, we will assume that $\Phi_0=\phi$. Proving conversely that this is the only solution of the nonlinear partial differential equations describing the leading force balance in the cross-section of the rod is beyond the scope of this analysis.

\vspace{-6pt}
\subsubsection*{\textbf{Solution at order $\vec{O(1)}$}}
At leading order, the incompressibility condition $\det{\tens{F}}=1$ reduces to~\footnote{Expansions were carried out using \textsc{Mathematica} to assist with manipulating complicated algebraic expressions.}
\begin{align}
\dfrac{Z_0Z_{0,\zeta}}{\zeta}=1\quad\Longrightarrow\quad Z_0=\zeta,    
\end{align}
using $Z_0=0$ on $\zeta=0$ by definition of the midsurface. At leading order, the Cauchy equation~\eqref{eq:cauchy} is
\begin{align}
\dfrac{\partial}{\partial \zeta}\bigl(\zeta\tens{P_0}\cdot\vec{N}\bigr)=\dfrac{\partial}{\partial \phi}\bigl(\tens{P_0}\cdot\vec{B}\bigr),
\end{align}
with
\begin{align}
&\tens{P_0}\cdot\vec{N}=S_{0,\zeta}\vec{\tilde{t}}+(1-p_0)\vec{\tilde{N}},&&\tens{P_0}\cdot\vec{B}=(1-p_0)\vec{\tilde{B}}.
\end{align}
Hence, recalling that $\vec{\tilde{N}}_{,\phi}=-\vec{\tilde{B}}+O(\varepsilon)$, $\vec{\tilde{B}}_{,\phi}=\vec{\tilde{N}}+O(\varepsilon)$ from definitions~\eqref{eq:BN} and $\tilde{\phi}_{,\phi)}=1+O(\varepsilon)$, and requiring that 
\begin{align}
(\zeta S_{0,\zeta})_{,\zeta}=0,&&\bigl(\zeta(1-p_0)\bigr)_{,\zeta}=(1-p_0),&&0=(1-p_0)_{,\phi},
\end{align}
which are to be solved subject to $S_0=0$ on $\zeta=0$ by definition of the midline, and the force-free conditions $\vec{P_0}\cdot\vec{N}=\vec{0}$, i.e. $S_{0,\zeta}=0$ and $p_0=1$, on $\zeta=h$. The solution of this leading-order problem is
\begin{align}
&S_0\equiv 0,&&p_0\equiv 1&&\Longrightarrow&&\tens{P_0}\equiv\tens{0}.
\end{align}
\subsubsection*{\textbf{Solution at order $\vec{O(\varepsilon)}$}}
We now turn to the solution at next order, at which the incompressibility condition $\det{\tens{F}}=1$ yields
\begin{align}
Z_{1,\zeta}+\Phi_{1,\phi}+\dfrac{Z_1}{\zeta}=-E+K\zeta\sin{\phi}.\label{eq:vc1}
\end{align}
Moreover, we compute
\begin{subequations}
\begin{align}
\tens{P_1}\cdot\vec{N}&=S_{1,\zeta}\vec{\tilde{t}}+(2Z_{1,\zeta}-p_1)\vec{\tilde{N}}-\left(\dfrac{Z_{1,\phi}}{\zeta}-\zeta\Phi_{1,\zeta}\right)\vec{\tilde{B}},\\
\tens{P_1}\cdot\vec{B}&=-\left(\dfrac{Z_{1,\phi}}{\zeta}\!-\!\zeta\Phi_{1,\zeta}\right)\vec{\tilde{N}}+\left(\dfrac{2Z_1}{\zeta}\!+\!2\Phi_{1,\phi}\!-\!p_1\right)\vec{\tilde{B}}.
\end{align}
\end{subequations}
Since $\tens{P_0}=\tens{0}$, the Cauchy equation at this order is
\begin{align}
\dfrac{\partial}{\partial \zeta}\bigl(\zeta\tens{P_1}\cdot\vec{N}\bigr)=\dfrac{\partial}{\partial \phi}\bigl(\tens{P_1}\cdot\vec{B}\bigr),
\end{align}
subject to $\tens{P_1}\cdot\vec{N}=\vec{0}$ on $\zeta=h$. The first component of this equation yields
\begin{align}
\dfrac{\partial}{\partial\zeta}(\zeta S_{1,\zeta})=0\quad\Longrightarrow\quad S_1\equiv 0,
\end{align}
on imposing $S_1=0$ on $\zeta=0$ and $S_{1,\zeta}=0$ on $\zeta=h$. The remaining components yield the coupled partial differential equations
\begin{subequations}\label{eq:fb1}
\begin{align}
\dfrac{\partial}{\partial \zeta}\bigl(\zeta(2Z_{1,\zeta}-p_1)\bigr)&=-\dfrac{\partial}{\partial\phi}\left(\dfrac{Z_{1,\phi}}{\zeta}-\zeta\Phi_{1,\zeta}\right)\nonumber\\&\quad+\left(\dfrac{2Z_1}{\zeta}\!+\!2\Phi_{1,\phi}\!-\!p_1\right),\\
-\dfrac{\partial}{\partial \zeta}\left(\zeta\left(\dfrac{Z_{1,\phi}}{\zeta}-\zeta\Phi_{1,\zeta}\right)\right)&=\dfrac{\partial}{\partial\phi}\left(\dfrac{2Z_1}{\zeta}+2\Phi_{1,\phi}-p_1\right)\nonumber\\&\quad+\left(\dfrac{Z_{1,\phi}}{\zeta}-\zeta\Phi_{1,\zeta}\right).
\end{align}
\end{subequations}
We seek the solution of Eqs.~\eqref{eq:vc1} and \eqref{eq:fb1} in the form 
\begin{subequations}
\begin{align}
Z_1(s,\zeta,\phi)&=a(s,\zeta)+b(s,\zeta)\sin{\phi}\\
\Phi_1(s,\zeta,\phi)&=c(s,\zeta)\cos{\phi}\\
p_1(s,\zeta,\phi)&=u(s,\zeta)+v(s,\zeta)\sin{\phi},
\end{align}
\end{subequations}
in which the unknown functions $a,b,c,u,v$ will be required to be regular at $\zeta=0$. Equations~\eqref{eq:vc1} and \eqref{eq:fb1} now yield\begin{subequations}
\begin{align}
&\zeta a'+a=-\zeta E,\label{eq:a}\\
&\zeta(b'-c)+b=\zeta^2K,\label{eq:c}\\
&\zeta^2(2a''-u')+2\zeta a'-2a=0,\label{eq:u}\\
&\zeta^2(2b''-c'-v')+2\zeta(b'+c)-3b=0,\label{eq:b}\\
&\zeta^3c''+3\zeta^2c'+\zeta(b'-2c-v)-3b=0,\label{eq:v}
\end{align}
\end{subequations}
wherein dashes denote differentiation with respect to $\zeta$. Imposing regularity, we can integrate Eq.~\eqref{eq:a} to obtain
$a=-\zeta E/2$ and then obtain $u(s,\zeta)=U(s)$ from Eq.~\eqref{eq:u}. Next, we solve Eq.~\eqref{eq:c} as a linear algebraic equation for $c$. The result turns Eq.~\eqref{eq:v} into a linear algebraic equation for $v$, and substituting both results into Eq.~\eqref{eq:b} yields ${3b'-3\zeta b''-6\zeta^2b'''-\zeta^3b''''=0}$, the regular solution of which that satisfies $b=0$ on $\zeta=0$ as required by the definition of the midsurface is $b(s,\zeta)=B(s)\zeta^2$. Imposing $\tens{P_1}\cdot\vec{N}=\vec{0}$ on $\zeta=h$ finally yields $U=-E$ and $B=K/4$, so that
\begin{align}
Z_1&=-\dfrac{E}{2}\zeta+\dfrac{K}{4}\zeta^2\sin{\phi},&&\Phi_1=-\dfrac{K}{4}\zeta\cos{\phi}.
\end{align}
We have also obtained $p_1=-E+K\zeta\sin{\phi}$, but it will turn out that this is not required for calculating the elastic energy at the next order.
\subsubsection*{\textbf{Asymptotic expansion of the elastic energy at order $\vec{O\bigl(\varepsilon^2\bigr)}$}}
It turns out that, to calculate the elastic energy of the rod at leading order, we will not need to solve for the deformation of the rod at order $O\bigl(\varepsilon^2\bigr)$. It suffices to note that the incompressibility condition at this order is
\begin{widetext}
\vspace{-10pt}
\begin{align}
Z_{2,\zeta}+Z_2/\zeta+\Phi_{2,\phi}&=\dfrac{3E^2}{4}+Eg+\dfrac{5K^2}{16}\zeta^2+\dfrac{\kappa K}{2}\zeta-\dfrac{K}{8}(5K+6\kappa)\zeta^2\cos{\phi}-\dfrac{E}{8}(7K+4\kappa)\zeta\sin{\phi}.
\end{align}

\vspace{-4pt}
\end{widetext}
Onto substituting this result into the expansion for the energy density in Eq.~\eqref{eq:E}, the latter reduces to
\begin{align}
e=\dfrac{C}{2}\left[3E^2-6EK\zeta\sin{\phi}+\dfrac{3K}{2}\zeta^2(1-\cos{2\phi})\right]\varepsilon^2+O\bigl(\varepsilon^3\bigr).
\end{align}
Since $\det{\tens{F^0}}=1+O(\varepsilon)$, this yields
\begin{subequations}
\begin{align}\label{eqn:rodenergywithzeta}
\mathcal{E}&=\varepsilon^4\!\!\int_{\mathcal{C}}{\left[\int_0^h{3\pi C\left(E^2+\dfrac{\zeta^2K^2}{2}\right)\zeta\,\mathrm{d}\zeta}\right]\mathrm{d}s}+O\bigl(\varepsilon^5\bigr),
\end{align}
where the outer integral is along the undeformed midline $\mathcal{C}$ of the rod. This expression for the dependence of the energy density on $\zeta$ will underlie our approximate calculations in the final section of this Supplemental Material. Performing the integration with respect to $\zeta$ finally yields
\begin{align}
\mathcal{E}=\varepsilon^2\bigl(\pi \varepsilon^2h^2\bigr)\int_{\mathcal{C}}{C'\left(E^2+\dfrac{h^2K^2}{4}\right)\,\mathrm{d}s}+O\bigl(\varepsilon^5\bigr),
\end{align}
\end{subequations}
with $C'=3C/2$, which, on recognising $\pi \varepsilon^2h^2$ as the cross-sectional area of the rod, is the asymptotic form of the energy density used in the introduction of the main text. In what follows, we will use the non-asymptotic form obtained by setting $\varepsilon=1$ or, more formally, replacing $\varepsilon h\mapsto h$, $\varepsilon E\mapsto E$, and we will also replace $C'\mapsto C$ for ease of notation, in agreement with the notation in the main text.
\section{Buckling of a uniformly growing rod}
We can now provide a more formal derivation of the buckling threshold of a uniformly growing rod that confirms the informal argument in the main text. This supports the use a similar argument in the calculation of the approximate buckling threshold for growth islands in the final section of this Supplemental Material. 

We assume a planar buckled midline $y(x)$, where ${x\in[-1/2,1/2]}$ is the coordinate along the straight undeformed midline of the rod. From definitions~\eqref{eq:EK}, the midline strain and curvature strain are
\begin{align}
E=\sqrt{1+y'(x)^2}-(1+g),&&K=\dfrac{y''(x)}{[1+y'(x)]^{3/2}}.
\end{align}
Varying $\mathcal{E}$ with respect to $y(x)$ leads to the Euler--Lagrange equation for the midline of the buckled rod,
\begin{align}
&\dfrac{h^2}{4[1+y'(x)^2]^3}\left(y''''(x)+\dfrac{6y''(x)^3}{[1+y'(x)^2]^2}-\dfrac{6y''(x)y'''(x)}{1+y'(x)^2}\right)\nonumber\\
&\quad=\left(\dfrac{1+g}{[1+y'(x)^2]^{3/2}}-1\right)y''(x),
\end{align}
subject to the boundary conditions ${y(\pm 1/2)=y'(\pm 1/2)=0}$ imposing the conditions of clamped, fixed ends. Anticipating a buckled solution of small amplitude $A\ll 1$, we expand ${y(x)=Ay_1(x)+O\bigl(A^2\bigr)}$, so that
\begin{align}
y_1''''(x)+\dfrac{4g}{h^2}y_1''(x)=0,\label{eq:ell}
\end{align}
subject to $y_1(\pm 1/2)=y_1'(\pm 1/2)=0$. The general solution of Eq.~\eqref{eq:ell} is
\begin{align}
y_1(x)=c_1+c_2x+c_3\cos{qx}+c_4\sin{qx},\label{eq:ysol}
\end{align}
where $q^2=4g/h^2$ and $c_1,c_2,c_3,c_4$ are constants of integration. The boundary conditions require
\begin{subequations}
\begin{align}
c_1+c_3\cos{\dfrac{\lambda}{2}}&=c_3\lambda\sin{\dfrac{\lambda}{2}}=0\label{eq:c1c3}\\
\dfrac{c_2}{2}+c_4\sin{\dfrac{\lambda}{2}}&=c_2+c_4\lambda\cos{\dfrac{\lambda}{2}}=0.\label{eq:c2c4}
\end{align}
\end{subequations}
In a \emph{symmetric} buckled mode, $c_2=c_4=0$, so ${q\sin{(q/2)}=0}$ from Eqs.~\eqref{eq:c1c3} unless $c_1=c_3=0$ and hence $y_1(x)\equiv 0$. The solution $q=0$ again leads to $y_1(x)\equiv 0$, so the lowest non-trivial eigenvalue, i.e., the smallest non-trivial wavenumber, is $q_\text{s}=2\pi$. \emph{Asymmetric} buckled modes with $c_2,c_4\not=0$ are also possible, but the simultaneous equations~\eqref{eq:c2c4} have a non-zero solution if and only if $\tan{(q/2)}=q/2$, with smallest non-trivial wavenumber $q_\text{a}\approx 8.99$. Thus $q_\text{a}>q_\text{s}$, so the symmetric mode is the dominant buckling mode.

By definition, $q^2=4g/h^2$, so the rod buckles when ${g=h^2 q_\text{s}^2/4=h^2\pi^2}$. With $q=q_\text{s}$, we also get ${c_1=c_3=1/2}$ up to rescaling $A$ from Eqs.~\eqref{eq:c1c3}. Equation~\eqref{eq:ysol} yields $y_1(x)=(1+\cos{2\pi x})/2=\cos^2{\pi x}$ because $c_2=c_4=0$ in a symmetric mode. Thus $y(x)=A\cos^2{\pi x}$ as assumed in the main text. 
\section{Numerical details}
\subsection{Finite-element model}
In our finite-element calculations, we consider a rod of unit length and circular cross-section of fixed radius $h=1/20$. We take cylindrical coordinates $(\zeta,\phi,x)$ in the undeformed rod, with $0\leq\zeta\leq h$, $0\leq\phi\leq 2\pi$, $-1/2\leq x\leq 1/2$.

Again, we describe the mechanics of the rod within the framework of morphoelasticity~\cite{rodriguez94,goriely,ambrosi19} and decompose the deformation gradient $\tens{\tilde{F}}$ into an elastic tensor $\tens{F}$ and a growth tensor $\tens{F_g}$, so that $\tens{\tilde{F}}=\tens{F}\tens{F_g}$, with
\begin{equation}
  \tens{F_g} =
  \begin{pmatrix}
  1+g &&& 0 &&&0 \\
  0 &&& 1 &&& 0 \\
  0 &&& 0 &&& 1
  \end{pmatrix}. 
  \label{eqn:Fg}
\end{equation}
for the purely axial growth considered here, where $g(x,\zeta,\phi)$ expresses the magnitude of this growth.

\begin{figure}[b]
\centering\includegraphics{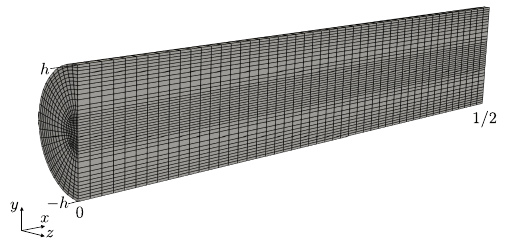}
\caption{Finite element mesh for the quarter rod $0\leq x\leq 1/2$, $-\pi/2\leq\phi\leq\pi/2$ plotted with respect to the Cartesian basis $(x,y,z)$). Parameter value: $h=1/20$.
\label{figS1}}
\end{figure}

Due to the axial and azimuthal symmetry of the growth assumed here [viz., $g(x,\zeta,\phi)=g(x,\zeta)=g(-x,\zeta)$], we can restrict the finite-element mesh for the numerical solution to the quarter rod $0\leq x\leq 1/2$, $-\pi/2 \leq \phi \leq\pi/2$. This mesh is shown in Fig. \ref{figS1} in the Cartesian basis $(x,y,z)$. On the two end faces (${x=0}$, ${x=1/2}$) and on the lateral face ($z=0$), we apply roller boundary conditions that constrain the displacement to be in the plane of these faces, respecting the symmetry of the problem.

To solve the buckling problem, we assume the simplest, neo-Hookean constitutive relations for the rod, so that its energy density \cite{holzapfel2000nonlinear,pelteret2016b} is
\begin{equation} \label{eqn:neohookean}
e = \frac{\mu}{2}\operatorname{tr}\bigl(\bar{\tens{F}}\bar{\tens{F}}^\top\bigr) + \frac{\kappa}{4}\bigl(J^2-1-2\log{J}\bigr),
\end{equation}
where $\bar{\tens{F}}=J^{-1/3}\tens{F}$, with $J=\det{\tens{F}}$, is the deviatoric part of the elastic tensor. Further, $\mu$ and $\kappa$ are the shear and bulk moduli related via Poisson's ratio $\nu$ by \cite{pelteret2016b}
\begin{equation}
    \kappa = \frac{2 \mu (1 + \nu)}{3 (1 - 2 \nu)},
\end{equation}
Because exact incompressibility, $J=1$, cannot be imposed in our numerical framework, we choose $\nu=0.499$ to make the rod nearly incompressible.

With these constitutive relations, the force balance equation for the rod is Cauchy's equation
\begin{align}\label{eqn:cauchystress}
\operatorname{div}{\tens{T}}=0,\qquad\text{where }\tens{T} = J^{-1} \tens{F} \frac{\partial e}{\partial \tens{F}^\top\tens{F}}\tens{F}^\top
\end{align}
is the Cauchy stress tensor. We solve this nonlinear problem using an iterative approach, with small increments of mean growth $\upDelta\langle g\rangle=3.5\times 10^{-5}$. At each growth increment, we solve equilibrium equations linearised around the deformed configuration of the previous step. This linearisation of the weak form of the equations and its spatial discretisation, and the solution algorithm are provided in Refs.~\cite{DealIITutorial2012,pelteret2016b}. The buckling threshold is obtained as the point of steepest increase, as growth is increased, of the displacement of the midpoint of the rod.

\subsection{Growth disorder}
To introduce growth disorder, we partition the rod into $M$ axial and $N$ radial regions in such a way that the resulting $MN$ cells have equal volumes. These cells are, in general, different from the elements of the mesh in Fig.~\ref{figS1}, with each cell containing multiple mesh elements. 

To each cell $c$ in this partition, we then assign a growth disorder $G_c$ such that the growth tensor takes the values 
\begin{equation}
  \tens{F}_{\tens{g},c} =
  \begin{pmatrix}
  1+\langle g \rangle (1+ G_c) &&& 0 &&&0 \\
  0 &&& 1 &&& 0 \\
  0 &&& 0 &&& 1
  \end{pmatrix}
  \label{eqn:Fg_disorder}
\end{equation}
on cell $c$, where $\langle g\rangle$ is the spatially averaged growth of the rod, which, because the cells have equal volumes, we impose by requiring
\begin{align}
\sum_c{G_c}=0.\label{eq:vc}
\end{align}
\subsubsection*{\textbf{Growth islands}}
In the case of ``growth islands'', the growth vanishes (i.e., ${G_c=-1}$) on all cells save the ones that represents the growth islands. Thus, from Eq.~\eqref{eq:vc},
\begin{align}
G_c=\left\{\begin{array}{cl}
-1&\text{if }c\notin\text{growth islands},\\
MN-1&\text{otherwise}.
\end{array}\right.
\end{align}
Varying $M,N$ allows sampling of different sizes of growth islands. Their position is varied by varying the choice of the cell corresponding to the growth island.

\subsubsection*{\textbf{Growth distributions}}
For the ``mechanical statistics'' in the main text, we generate random growth fields by sampling $MN$ random numbers $G_c$ from the uniform distribution on the $MN$-dimensional hypercube with sides $[-a,b]$ subject to constraint~\eqref{eq:vc} using the \texttt{RandFixedSum} algorithm~\cite{rfs} implemented in \textsc{Matlab} (The MathWorks, Inc.). Here, $a,b>0$ and $a\leq 1$ so that $g_c=\langle g\rangle (1+G_c)\geq 0$ for each cell $c$.

In our statistical analysis in Fig.~3 of the main text, we use $(M,N)=(30,1)$ [case (1)], $(M,N)=(1,12)$ [case~(2)], and $(M,N)=(30,12)$ [case (3)]. We sample configurations of small variance by taking $b=a\leq 1$ and configurations of larger variance by taking $a=1,b>1$.

\section{Buckling of a rod with growth islands}
In this section, we provide the details of the derivation of the approximation, stated in the main text, of the buckling threshold of a rod in which the growth is concentrated in ``growth islands'', as introduced in the main text. The growth islands [Fig. 2(d) of the main text] are two symmetric annular segments, each of length $\ell_{\text{g}}/2$ and cross-sectional area $\smash{\pi h_{\text{g}}^2}$, centred at $x=\pm x_0$ and $\zeta=\zeta_0$ and defined by
\begin{align}
  \pm x_0-\frac{\ell_\mathrm{g}}{4}  \leq x \leq \pm x_0+\frac{\ell_\mathrm{g}}{4},
    &&\zeta_0^2-\frac{h_\mathrm{g}^2}{2h^2} \leq \zeta^2 \leq \zeta_0^2+\frac{h_\mathrm{g}^2}{2h^2},\label{eqn:bound}
\end{align}
The growth islands thus occupy a fraction $\chi_\mathrm{g}={\ell_\mathrm{g} h^2_\mathrm{g}}/{h^2}$ of the total volume of the rod. The growth field is therefore
\begin{align}
g(x,\zeta)=\left\{\begin{array}{cl}
\vspace{1.5mm}
\dfrac{\langle g\rangle}{\chi_\mathrm{g}}&\text{if }(x,\zeta)\in\text{growth islands},\\
0&\text{otherwise},
\end{array}\right.
\end{align}
where $\langle g\rangle$ is the (spatially averaged) mean growth of the rod. Indeed, the quantity ${\chi_\mathrm{g}(1\!+\!\langle g\rangle/\chi_\mathrm{g})\!+(1\!-\!\chi_\mathrm{g})(1\!+\!0)=1\!+\!\langle g\rangle}$ is then the total growth of the rod, as required.

Due to the expected symmetry of the lowest buckling mode, we can, similarly to the numerical solution, restrict our calculations to the half rod $0\leq x\leq 1/2$. 

We now make our first assumption: We assume that the geometric stretch $\lambda(x,\zeta)$ in the flat configuration of the rod is piecewise uniform along the rod and uniform in the radial direction. This is consistent with the simple elastic stretch fields observed in numerical solutions in cases (1) and~(2) in Fig. 2(b) of the main text, but is clearly a simplification in case~(3). We thus assume
\begin{align}
\lambda(x,\zeta)=\left\{\begin{array}{cl}
\lambda_1&\text{if }x_0-\dfrac{\ell_\mathrm{g}}{4}  \leq x \leq x_0+\dfrac{\ell_\mathrm{g}}{4},\\
\lambda_2&\text{otherwise},
\end{array}\right.
\end{align}
where the constants $\lambda_1,\lambda_2$ are to be determined in our calculation. In the flat configuration, the condition of fixed ends imposes
\begin{equation}\label{eqn:consistency}
  \frac{\ell_\mathrm{g}}{2} \lambda_1+ \left(\frac{1}{2}-\frac{\ell_\mathrm{g}}{2}\right) \lambda_2 = \frac{1}{2}.
\end{equation}
We further assume that the deflection of the rod, in its buckled configuration, is still given by that of the uniform rod, viz.,
\begin{align}
y(x)=A\hat{y}(x)+O\bigl(A^2\bigr),\quad \text{where }\hat{y}(x)=\cos^2{\pi x}.
\end{align}
Let $\lambda_{\text{g}}=1+g$ denote the intrinsic growth stretch. The elastic stretch of the midline in the buckled configuration is then, from the multiplicative decomposition of morphoelasticity,
\begin{align}
\Lambda=\dfrac{\sqrt{\lambda^2+y'(x)^2}}{\lambda_{\text{g}}}=\dfrac{\lambda}{\lambda_{\text{g}}}+A^2\dfrac{\hat{y}'^2}{2\lambda\lambda_{\text{g}}}+O\bigl(A^4\bigr).
\end{align}
Because we seek to compute a correction to the leading-order buckling threshold, we need to replace the strain energy term $E^2$ in the energy density~\eqref{eqn:rodenergywithzeta} with a nonlinear form. We choose the neo-Hookean energy density corresponding to principal stretches $\Lambda,\Lambda^{-1/2},\Lambda^{-1/2}$, as required by volume conservation, for this purpose and thus replace
\begin{align}
E^2&\to \Lambda^2+\dfrac{2}{\Lambda}-3\nonumber\\
&\qquad=\left(\dfrac{\lambda}{\lambda_{\text{g}}}\right)^2+\dfrac{2\lambda_{\text{g}}}{\lambda}-3+A^2\dfrac{\hat{y}'(x)^2}{\lambda\lambda_{\text{g}}}\dfrac{\left(\lambda/\lambda_{\text{g}}\right)^3\!-\!1}{\left(\lambda/\lambda_{\text{g}}\right)^2}+O\bigl(A^4\bigr)\nonumber\\
&\qquad\approx \left(\dfrac{\lambda}{\lambda_{\text{g}}}\right)^2+\dfrac{2\lambda_{\text{g}}}{\lambda}-3+A^2\hat{y}'(x)^2\dfrac{\lambda-\lambda_{\text{g}}}{\lambda}.
\end{align}
In the final step, we have linearised, in the term at order $O\bigl(A^2\bigr)$, $\lambda\approx\lambda_{\text{g}}\approx 1$ excepting one factor $1/\lambda$ resulting from the midline stretch, and, of course, except in the difference $\lambda-\lambda_{\text{g}}$. We emphasise that we have no asymptotic justification for the use of this nonlinear midline strain in our calculation~\footnote{This justification would be provided by extending the asymptotic derivation of the rod theory in the first section of this Supplemental Material to non-uniform growth and beyond leading order, which is beyond the scope of this analysis.}. Moreover, the curvature strain of the buckled rod is
\begin{widetext}

\vspace{-14pt}
\begin{align}
K=\dfrac{\lambda y''(x)}{\bigl(\lambda^2+y'(x)^2\bigr)^{3/2}}\approx A\dfrac{\hat{y}''(x)}{\lambda^2},
\end{align}
\noindent which includes a factor $1/\lambda^2$ resulting from the stretch of the midline; this provides an informal justification for not neglecting a corresponding factor $1/\lambda$ above. Using Eq.~\eqref{eqn:rodenergywithzeta}, we thus obtain an approximation for the energy of the rod,
\begin{align}
\mathcal{E}\approx\int_0^1{\int_0^h{\left[\left(\dfrac{\lambda}{\lambda_{\text{g}}}\right)^2+\dfrac{2\lambda_{\text{g}}}{\lambda}-3+A^2\left(\hat{y}'(x)^2\dfrac{\lambda-\lambda_{\text{g}}}{\lambda}+\dfrac{\zeta^2\hat{y}''(x)^2}{2\lambda^4}\right)\right]\lambda_{\text{g}}\,\mathrm{d}\zeta}\,\mathrm{d}x}=\mathcal{E}_0+A^2\mathcal{E}_2.
\end{align}
In the double integral, we have included a factor of $\lambda_{\text{g}}\approx 1$ that corresponds to the Jacobian for the transformation from the undeformed to the intrinsic configuration of the rod, motivated, like the use of the nonlinear form for the strain energy, by our quest to compute an estimate for a correction to the buckling threshold. Again, we do not have a formal asymptotic justification for including this term. In the final expression, $\mathcal{E}_0,\mathcal{E}_2$ are complicated functions of $\lambda_1,\lambda_2,\langle g\rangle_\ast$ as well as of the geometric parameters describing the growth islands. We determine $\lambda_1,\lambda_2$ by minimising $\mathcal{E}_0$ subject to Eq.~\eqref{eqn:consistency}. For this purpose, we write
\begin{align} \label{eq:gapprox}
\lambda_1&=1+L_2h^2+L_4h^4+O\bigl(h^6\bigr),&\langle g\rangle_\ast &=g_2h^2+g_4h^4+O\bigl(h^6\bigr),
\end{align}
which yields an expansion for $\lambda_2$ from Eq.~\eqref{eqn:consistency}. We thus obtain 
\begin{align}
L_2&=-g_2\left(\dfrac{1}{\ell_{\mathrm{g}}}-1\right),&&L_4=\bigl(g_2^2-g_4\bigr)\left(\dfrac{1}{\ell_{\mathrm{g}}}-1\right).
\end{align}
The buckling threshold is now determined from $\mathcal{E}_2=0$, which corresponds to $\partial\mathcal{E}/\partial A^2=0$ at $A=0$ [Fig. 1(c) of the main text], and is therefore when the energy of buckled shapes ($A\neq0$) becomes lower than that of the flat configuration ($A=0$). We find
\begin{align}
g_2&=\pi^2,&&g_4=\dfrac{\pi^3}{\ell_{\text{g}}^2}\left\{\pi\ell_{\text{g}}\left[1-\ell_{\text{g}}+2\ell_{\text{g}}\left(\dfrac{\zeta_0}{h}\right)^2-\left(\dfrac{h_{\text{g}}}{h}\right)^{-2}\right]-\left[1+3\ell_{\text{g}}-2\ell_{\text{g}}\left(\dfrac{\zeta_0}{h}\right)^2-\left(\dfrac{h_{\text{g}}}{h}\right)^{-2}\right]\cos{\bigl(4\pi x_0\bigr)}\sin{\bigl(\pi\ell_{\text{g}}\bigr)}\right\}.
\end{align}
\end{widetext}
In particular, $\langle g\rangle=\pi^2h^2+O\bigl(h^4\bigr)$, consistently with the asymptotic result for uniform growth. Moreover, for uniform growth, $\ell_{\text{g}}=1,h_{\mathrm{g}}=1,x_0=1/4,\zeta_0=\smash{h/\sqrt{2}}$, and this approximation yields $g_4=0$, and hence, from the second of Eqs.~\eqref{eq:gapprox}, $\smash{g_\ast=\pi^2h^2+O\bigl(h^6\bigr)}$. Equation (3) of the main text then follows. Equation (1) of the main text follows as a special case, by taking $h_{\text{g}}=h$, $\smash{\zeta_0=h/\sqrt{2}}$, corresponding to the growth islands spanning the thickness of the rod. Equation (2) follows similarly by taking $\ell_{\mathrm{g}}=1/2$, $x_{\mathrm{g}}=1/4$, for which the growth island spans the entire length of the rod. 

\bibliography{main.bib}